\documentclass[paper]{JHEP3}

\usepackage{graphicx}
\usepackage{amsmath}
\usepackage{amsfonts}
\usepackage{amscd}
\usepackage{epsfig}
\usepackage{amssymb}
\usepackage{tabularx}
\usepackage{amsthm} 
\newtheorem{theorem}{Theorem}

\def\c{{\mathbb C}}

\def\a{\alpha}
\def\b{\beta}

\def\e{\epsilon}

\def\L{\Lambda}
\def\m{{\cal M}}
\def\mc{{\cal M}_{c}}
\def\hp{x}

\def\ssqr#1#2{{\vbox{\hrule height #2pt
\hbox{\vrule width #2pt height#1pt \kern#1pt\vrule width #2pt}
\hrule height #2pt}\kern- #2pt}}

\newcommand{\drawsquare}[2]{\hbox{%
\rule{#2pt}{#1pt}\hskip-#2pt%  left vertical
\rule{#1pt}{#2pt}\hskip-#1pt%  lower horizontal
\rule[#1pt]{#1pt}{#2pt}}\rule[#1pt]{#2pt}{#2pt}\hskip-#2pt%  upper horizontal
\rule{#2pt}{#1pt}}% right vertical
 
\newcommand{\Yfund}{\raisebox{-.5pt}{\drawsquare{6.5}{0.4}}}%  fund
\newcommand{\Yasymm}{\raisebox{-3.5pt}{\drawsquare{6.5}{0.4}}\hskip-6.9pt%
        \raisebox{3pt}{\drawsquare{6.5}{0.4}}}%  antisymmetric second rank
\newcommand{\Ythreea}{\raisebox{-3.5pt}{\drawsquare{6.5}{0.4}}\hskip-6.9pt%
        \raisebox{3pt}{\drawsquare{6.5}{0.4}}\hskip-6.9pt
        \raisebox{9.5pt}{\drawsquare{6.5}{0.4}}}           

\title{Topology of Quantum Modified Moduli Spaces}
\author{Gustavo Dotti \\FaMAF, Universidad Nacional de C\'ordoba, Ciudad Universitaria, 
(5000) C\'ordoba, Argentina \\
E-mail: \email{gdotti@fis.uncor.edu}}

\abstract{
We prove that all SYM theories 
that have  a quantum modified moduli space   $\m$ defined by 
a single constraint equation have trivial homotopy groups 
$\pi_j(\m)$ for $j=0,1,2,3$ and $4$. This implies that 
none of these theories  admit skyrmions or vortexes
-a fact that had  only been proved 
for supersymmetric QCD with $N_f=N_c$ and $Sp(2N)$ 
with $2N+2$ fundamentals- 
whereas those of them with a nontrivial 
$H^5 (\m)$  admit  Wess-Zumino-Witten terms in their 
effective actions. Contrary to expectations, examples of 
quantum modified moduli spaces with a trivial $H^5 (\m)$
are found in the literature. 
}

\keywords{set, dag}

\preprint{hep-th/0306264}

\begin{document}

\section{Introduction}

The set of supersymmetric vacua modulo gauge transformation of a SYM theory
is known as the {\em moduli space}. The {\em classical} moduli space $\mc$ 
is parametrized by a basic set of holomorphic gauge invariant operators $\hp_i(\phi), 
i=1,...,n$, $\phi$ the elementary chiral matter fields.
 Generically, the $\hp's$ are subjected to polynomial constraints 
$c_a(\hp(\phi)) \equiv 0, a=1,...,r$,  then $\mc \subset \c^n$ is 
the algebraic set $\{ x \in \c^n |\; c_a(x)=0,  a=1,...,r \}$, 
the zero set of $r$ 
polynomials in $n$ complex variables \cite{plb,t}. 
For  those SYM theories having  multiple quantum  supersymmetric vacua,
the  {\em quantum} moduli space $\m$ is parametrized by the {\em vevs} $<x_i>$ 
of the basic invariants, and $\m$ is also an algebraic subset of $\c^n$. 
In fact either $\m$ or a branch of it equals  $\mc$ for most theories. 
However, if the matter content of a theory is in a 
 gauge group 
representation  whose     Dynkin index $\mu_{\rho}$ equals   the 
adjoint index $\mu_{adj}$, and the theory 
does not have 
D-flat points that break the gauge group to $U(1)^k$, 
then one of the constraints that define $\mc$, say 
$c_r$, gets
quantum modified to either $c_r(\hp) = \Lambda^p$ or 
$c_r(\hp) = \hp_k \Lambda^p$, smoothing out  the singularities of 
the affine variety $\mc$ \cite{sei,susy,gn1,gn2}. These are the theories with 
a quantum modified moduli space (QMMS), of which those defined 
by a single constraint ($r=1$) are the subject of this paper.
 All classical moduli spaces are contractible, and then homotopically trivial.
The QMMS, instead, 
are suspected of being smooth complex manifolds 
of a nontrivial homotopy type, although these facts have only been proved 
for SQCD with equal number of colors and flavors, and for
$SP(2N)$ with $2N+2$ fundamentals \cite{m,r}. 
The interest in the topology of $\m$ is due to the fact 
that, if  nontrivial, some special 
(topological) terms can be added to the 
effective action of the theory. Also, topological stable field configurations such as Skyrmions or 
vortexes may be possible. 
Among the  interesting topological 
invariants  are   $\pi_2(\m)$, in connection 
with the existence of vortexes \cite{wzw}, and  $\pi_3(\m)$, which, if nontrivial, implies 
the existence of skyrmions \cite{sky,wzw}. Also, if $\pi_4(\m)$ is trivial 
and $H_5(\m,{\mathbb R})$ is 
non trivial, the theory admits Wess-Zumino-Witten terms in its 
effective action \cite{wz,wzw}.
The calculation of these homotopy and cohomology groups looks like a
formidable task at first sight, since the $\m$'s are  complex algebraic 
sets  defined by polynomials in a large number of variables.
Non supersymmetric gauge theories usually exhibit 
a single $H$ orbit of vacua, $H$ the global symmetry group. Their 
moduli spaces are then homogeneous spaces $H/H_o$,  
whose topology is well known. 
In the supersymmetric case, 
however, $\m$ is a non compact space where $H$ does not act 
transitively. In fact $\m$ contains strata of $H$ orbits of different 
kinds. The only exceptions are $SP(2N)$ with 
$(2N+2)\Yfund$ matter fields  and 
 SQCD with $N_c=N_F \equiv N$, in the special case 
$N=2$ \cite{m}. 
The reason why $N=2$  SQCD is special is that 
 the fundamental and antifundamental of $SU(N)$ 
are equivalent when $N=2$, and so the flavor group, which is  
$SU(N) \times SU(N) \times U(1)$ if $N>2$, gets enlarged to $SU(4)$ when 
$N=2$. It is under this larger flavor group that  the quantum moduli space 
of SQCD with  $N=2$ becomes a single orbit, and thus 
a homogeneous space. The fact that the moduli spaces of 
SQCD with two colors and flavors, and  $SP(2N)$ with 
$(2N+2)\Yfund$ matter fields are homogeneous 
 is what allowed the computation of their 
homotopy and cohomology groups in  \cite{m}. 
Unfortunately, it does not seem to be possible 
to apply this idea to other QMMS by, for example,  extending 
  $H$ to a (possibly anomalous) larger symmetry 
group $H'$ chosen to act  transitively on $\m$.
This may explain why the only other 
 available calculation of the homotopy type 
of a QMMS, SQCD with $N_c=N_f \geq 2$, 
performed in \cite{r}, uses a completely different 
technique, which,  however,  cannot be extrapolated to other 
examples either,  because it  strongly depends on  the 
specific form of the SQCD constraint.\\
In this paper we prove that  some relevant topological facts of the 
$N=2$ SQCD and $Sp(2N)$ with $(2N+2) \Yfund$ QMMS
 hold 
generically for  QMMS defined by a single constraint equation, 
of SYM theories based on  simple gauge groups. 
These theories are listed  in  \cite{gn1,gn2}, where they were classified  
into  two broad classes: (i) {\em invariant} QMMS, defined by $p(\hp)=\Lambda^d$, 
$p(\hp)$ a flavor singlet polynomial of mass dimension $d$ and  (ii) 
{\em covariant} QMMS, defined by $p(\hp)= \hp_k \Lambda^{d-d_k}$, 
$p(\hp)$ a dimension $d$ operator carrying a flavor $U(1)$ charge
equal to that of  $x_k$, an invariant of mass dimension $d_k$. 
In both cases the classical moduli space is the set defined by 
$p(\hp)=0$. This fact, together with some particular aspects of the stratification 
of the  classical moduli spaces of theories with a QMMS,  allowed us 
to prove that both types of QMMS 
have trivial $\pi_j(\m)$ for $j=0,1,2,3,4$, i.e., they are $4-${\em connected}.
Here we use the standard convention that $\pi_0(\m)$ is the set of connected 
components of $\m$. $\pi_0$ is not a group, the triviality of 
$\pi_0(\m)$ merely means that $\m$ is a connected set.\\
The paper is organized as follows: in Section \ref{strat} we 
review some fundamental aspects of the stratification of the 
classical moduli space of a SYM theory according to the unbroken gauge
subgroup at  different vacua, and also study the Higgs flows among theories with a
 QMMS.
The required stratification results, due to Luna, Procesi and  Schwarz, 
\cite{plb,l,s}, are collected in Theorem \ref{teo1}.  
A SYM theory is represented 
$[G,\rho]$, $G$ the gauge group, $\rho$ the $G$ representation
 of the matter content. Higgs flow is indicated $[G,\rho] \to 
[G',\rho']$, or just $G \to G'$ when the matter content is irrelevant.
 Theorem \ref{d1}, proved in this section, 
 states  that all theories with a QMMS flow to $[SU(2),4 \Yfund 
+ singlets]$. In Section \ref{alg} we re-derive the 
results in \cite{m,r} on the topology of ${\cal M}_{SQCD}$
using alternative techniques.
One of the derivations uses very 
recent algebraic geometric results due to Dimca and 
Paunescu \cite{d}, that we introduce as 
Theorem \ref{d&p}. 
In Section \ref{teo} we prove our main result, Theorem \ref{d2},
 which states that 
all QMMS defined by a single equation are $4$-connected. The proof 
uses the three previous theorems.
Section \ref{conc} contains the conclusions. 
For quick reference, we have gathered in a brief appendix 
a number of useful algebraic topology definitions and 
theorems.

\section{Stratification of the classical moduli space} \label{strat}

 We recall some facts about the classical moduli 
space of a supersymmetric gauge theory \cite{plb}.
$\phi \in \c^q = \{\phi \}$ denotes a spacetime constant
configuration of the elementary matter chiral fields.
 $G$ is the gauge group, $\rho$ its representation
 on $\{ \phi \}$,  $\rho = \oplus_{i=1}^k F_i \rho_i$ its decomposition
into irreducible representations.
$\hp_{i}(\phi), i=1,...,n$ is a basic set of homogeneous, holomorphic
$G$ invariant polynomials  on $\c^q$. 
The invariants are subjected to polynomial constraints 
$c_a(\hp(\phi)) \equiv 0, a=1,...,r$. 
There is precisely one $G$ orbit of $D-$flat points in 
every {\em fiber} $\{ \phi \in \c^q | \hp(\phi)=\hp_o,  \;\;c_a(\hp_o)=0 \}$
\cite{plb}, then,
for theories with zero
superpotential,
the classical moduli space $\mc \equiv  \{D-\text{flat points}\}/G$ 
equals the set 
$\{ \hp \in \c^n |  \; c_a(\hp) = 0, a=1,...,r \} = \hp(\c^q) \subseteq \c^n$.             
Given $g \in G$, the isotropy subgroups at  $\phi$ and $g \phi$ are 
conjugated: $G_ {g \phi} = g G_{\phi} g^{-1}$. Since there is precisely 
one $G$ orbit of D-flat points in the fiber $\hp(\phi)=\hp_o$,
 a conjugacy 
class $(G_{\hp_o})$ can be associated to  $\hp_o \in \mc$. 
The stratum $\Sigma_{(H)} \subset \mc $ is defined 
by $\Sigma_{(H)} = \{ \hp \in \mc | (G_{\hp})=(H) \}$, i.e., 
two points of $\mc$ lie in the same stratum 
if their associated D-flat points have conjugate isotropy subgroups. 
$\mc$ is the disjoint union of its strata.
We will say $(H_1) \leq (H_2)$ if $H_1$ is conjugated 
to a subgroup of $H_2$. This is a {\em partial} order relation, 
given two classes, it may well happen that neither $(H_1) \leq (H_2)$
nor $(H_2) \leq (H_1)$ (see \cite{ls} for examples). 
There is a unique minimal class $(G_P)$,  
and certainly a unique maximal class, namely  $(G)$. $\Sigma_{(G_P)}$ is 
called the {\em principal stratum.} 
The vacua at $\Sigma_{(H)}$ correspond to D-flat points 
that break $G$ to a subgroup  $H$, $G_P$ being  the maximally broken 
subgroup of $G$. It can be shown that  $(H_2) \leq (H_1)$ if 
and only if it is possible to flow 
by Higgs mechanism from the $H_1$ gauge theory to the 
$H_2$ one.
$(H)$ is said to be {\em subprincipal} if its minimal among 
non principal classes. In general, there will be many subprincipal 
classes. 
A number of useful results related to the stratification of $\mc$, 
due to Luna and Schwarz,  are collected in the theorem below:
\begin{theorem} [Luna, Schwarz] \label{teo1} \cite{plb,l,s} 
\begin{enumerate}
\item There are only finitely many strata. The strata are smooth complex 
manifolds, whose closures are irreducible algebraic subsets of $\mc$
\item The closure of the  stratum of the class $(H)$ equals the union of the 
strata of greater or equal classes.
$$\overline{\Sigma_{(H)}} = \bigcup_{(L) \geq (H)} \Sigma_{(L)} $$
\item If $\hp$ is a singular point of $\mc$ then $\hp \notin \Sigma_{(H_P)}$
\item Consider Higgs mechanism at the D-flat point $\phi  \in \c^q$.
Let $N_{\phi}$ be  a $G_{\phi}$ invariant complement 
to the eaten field space  $Lie \left( G \right)  \phi$, 
$N_{\phi} = \rho_{\phi} \oplus s {\mathbb I}$ its
 decomposition  into $G_{\phi}$ singlets and non singlets, then 
\begin{equation} \label{is}
\c^q = Lie \left( {G_{\phi}}\right)  \phi \oplus  \rho_{\phi} \oplus s {\mathbb I}
\end{equation}
The restriction of the map $x: \c^q \to \mc$ to the singlet subspace 
$ s {\mathbb I}$  is a local coordinate 
chart $\hp:  s {\mathbb I} \to \Sigma_{(G_{\phi})}$   for the stratum.
In particular, the dimension of the stratum 
equals $s$, the number of singlets.
\end{enumerate}
\end{theorem}
Recall that an {\em algebraic set} $X$   is the set of zeroes  of a finite 
set of polynomials and  is said to be {\em irreducible} if 
it is not  the proper union of two algebraic sets. There is a notion 
of tangent space at $x \in X$, and $x$ is said to be a {\em singular point}
of $X$ if the dimension of the tangent at $x$ is different from the 
dimension of $X$ (see, e.g.,  \cite{cls}). Point 3 in the theorem 
states the well known fact that 
singular points of $\mc$ correspond to vacua with 
enhanced gauge symmetry. In  eqn(\ref{is}), 
$[G_{\phi},  \rho_{\phi} + s {\mathbb I}]$ is the theory towards which 
the original theory  $[G,\rho]$ flows
 by Higgs mechanism at the vacuum $\phi$, 
$Lie \left( G \right)  \phi$ being the eaten fields. 
We rarely keep track of the leftover singlets $ s {\mathbb I}$,  
because they are dynamically irrelevant. In what follows, however, we 
will need to know the dimensions of certain strata, which, 
according to 
Theorem 1.4, equal the number of singlets.\\
It is useful to display isotropy 
classes in decreasing order from left to right, with ordered strata 
connected by a line.
The resulting diagram encodes   all patterns of gauge symmetry breaking.\\
As an example, the diagram
\begin{equation} \label{ex}
\begin{array}{lclclcl}
& & (G_1) & - & (G_5) & \\
& \diagup &&&&\diagdown\\
(G) & - & (G_2) &- &(G_4) &- & (G_P)\\
& \diagdown & & \diagup & \\
&& (G_3) &&&
\end{array}
\end{equation}
 tells us that the sequence  of Higgs flows 
 $G \to G_3 \to G_4 \to G_P$ is possible 
(since $(G) \geq (G_3) \geq (G_4) \geq (G_P))$,
  whereas the sequence 
 $G \to G_2 \to G_5 \to G_P$ is not (since $(G_2) \ngeq (G_5)$).
In this example there are two subprincipal classes, $(G_4)$ and $(G_5)$. 
According to Theorem 1.2
\begin{equation} \label{ex1}
\overline{\Sigma_{(G_5)}}  = \Sigma_{(G_5)} \cup \Sigma_{(G_1)} 
\cup \Sigma_{(G)}, 
\hspace{1cm} 
\overline{\Sigma_{(G_4)}}  = \Sigma_{(G_4)} \cup 
\Sigma_{(G_2)} \cup \Sigma_{(G_3)} \cup \Sigma_{(G)}, 
\end{equation}
therefore, from Theorem 1.3, if $\phi$ is a singular point of $\mc$ 
\begin{equation} \label{ex2}
\phi \in \mc - \Sigma_{(G_P)} = \overline{\Sigma_{(G_5)}} \cup 
\overline{\Sigma_{(G_4)}}.
\end{equation}
Theories with a QMMS flow among themselves, and have trivial $G_P$.
The following theorem shows  that any 
sequence of Higgs flows from the QMMS  theory $[G,\rho]$ to 
the trivial theory $[1,singlets]$ has the form $[G,\rho] \to 
\cdots \to [SU(2),4 \Yfund + singlets] \to [1,singlets]$.
Surprisingly, this fact will turn out to 
be relevant to the computation of topological 
invariants of the quantum modified moduli spaces $\m$ of these theories. 

\begin{theorem} \label{d1}
If $[G,\rho]$ is a theory with a QMMS and $[G',\rho']$ 
is  a subprincipal stratum, then $G'=SU(2)$ and $\rho'=4 \Yfund + s {\mathbb I}$
\end{theorem}

\begin{proof}
 By definition of  subprincipal stratum, 
$[G',\rho']$ can only flow to a trivial theory by Higgs mechanism. 
Since $[G',\rho']$ is a QMMS theory, the theorem can be re-stated as
follows: 
if  every non zero 
D-flat point of the QMMS theory $[G',\rho']$ completely breaks $G'$, then 
$G'=SU(2)$ and $\rho'=4 \Yfund + s {\mathbb I}$. 
Note that every non zero D-flat point of $[SU(2),4 \Yfund + s {\mathbb I}]$ 
does break $SU(2)$ completely. Note also  that 
$G'$ cannot contain $U(1)$ factors,  
it must either be simple or semisimple. We will consider separately both cases.\\ 
If $G'$ is simple, then $[G',\rho']$ is among 
the QMMS theories listed in 
\cite{gn1,gn2}. With the exception of $[SU(2),4 \Yfund]$, which 
can only flow to a trivial theory, 
every one of  these theories flows by Higgs mechanism  to another 
QMMS theory,  with a simple or  semisimple gauge group (most of 
the flows involving $SU$ and $Sp$ theories are given in \cite{gn1}, 
we have calculated the remaining ones.)
This means that the only theory based  on a simple gauge group 
that can be a subprincipal stratum of a larger theory is 
$[SU(2),4 \Yfund + s {\mathbb I}]$. \\
Assume now  that $G'$ is semisimple, and 
for simplicity, that it  
contains only two simple factors
$G'=G_{(1)} \times G_{(2)}$ (our arguments generalize easily to the case  
where $G'$ contains more factors.)
 Let 
\begin{equation} \label{dec1}
\rho' = \sum_{i \a} c_{i \a} (\rho^{(1)}_i,\rho^{(2)}_{\a}) 
\end{equation}
be the decomposition of $\rho'$ into irreducible representations (irreps), 
$\rho^{(1)}_i$ a set of irreps of $G_{(1)}$, $\rho^{(2)}_{\a}$ irreps of $G_{(2)}$.
As a $G_{(1)}$ SYM theory,  $[G_{(1)} \times G_{(2)}, \rho']$ has  matter content
\begin{equation} \label{dec2}
\rho^{(1)} = \sum_{i} \left( \sum_{\a} c_{i \a} 
d_{\a} \right)  \rho^{(1)}_i \; , 
\hspace{.5cm}  d_{\a} =  \text{dim }(\rho^{(2)}_{\a}).
\end{equation}
 The theory $[G_{(1)},\rho^{(1)}]$ 
satisfies the index constraint $\mu = \mu_{adj}$ and 
does not flow to a $U(1)$ gauge theory, then it is a QMMS theory 
based on a simple gauge group, and so is listed in \cite{gn1,gn2}.
We will call this  theory the $G_{(1)}$ {\em projection} of 
(\ref{dec1}).\\
An inspection of the tables in \cite{gn1,gn2} shows that, with the 
exception of $[Sp(2n),(2n+2) \Yfund]$ and $[SU(2),4 \Yfund]$, 
all QMMS theories are of the form $[G, \sum_i f_i \rho_i]$ with 
$f_i \leq \text{ dim } \rho_i \equiv d_i$, i.e., the number of flavors 
of an irrep is less that or equal to its dimension. For later 
use, we have gathered in 
Table~ I
 below all theories having an $f_i \geq d_i - 2$. 
If we assume a semisimple subprincipal stratum such that   
neither the $G_{(1)}$ projection nor the $G_{(2)}$ projection of (\ref{dec1}) 
gives  $[Sp(2n),(2n+2) \Yfund]$ or $[SU(2),4 \Yfund]$, then the number 
of flavors of a given irrep in a projection never exceeds 
its dimension, and, according to 
(\ref{dec2}), if $c_{i \a} \neq0$ it must be 
$$ f_i \geq d_{\a} \geq f_{\a}$$
and also 
$$ f_{\a} \geq d_{i} \geq f_{i}.$$
This implies 
\begin{equation} \label{fla}
f_i \leq d_i \leq f_{\a} \leq d_{\a} \leq f_{i}
\end{equation}
and so all these numbers must be equal, meaning that 
the projections of (\ref{dec1}) must be among entries 
3 and 4 of table~ I. This leaves us with the following  possibilities:
\begin{equation}\label{pos1}
\begin{array}{ll}
SU(N) \times SU(N) & (\Yfund,\Yfund) + (\bar \Yfund,\bar \Yfund) \\
                                   & (\Yfund,\bar \Yfund) + (\bar \Yfund, \Yfund) \\
                     &  (\Yfund,\Yfund) + N(\bar \Yfund,1) + N (1, \bar \Yfund)\\
SU(4) \times Sp(4) & (\Yfund, \Yfund) + (1, \Yasymm) + 4 (\bar \Yfund, 1)\\
Sp(4) \times Sp(4) & (\Yfund, \Yfund) + (\Yasymm,1) + (1, \Yasymm)
\end{array}
\end{equation}
Theories containing an $SP(4)$ factor above 
flow to a QMMS theory with a semisimple gauge group  by a vev 
$\langle (1, \Yasymm) \rangle$, 
$[SU(N) \times SU(N), (\Yfund,\Yfund) +
 N(\bar \Yfund,1) + N (1, \bar \Yfund)]$ flows to SQCD 
by a vev $\langle  N (1, \bar \Yfund) \rangle$, 
$[SU(N) \times SU(N) ,  (\Yfund,\Yfund) + (\bar \Yfund,\bar \Yfund)]$
flows to a diagonal $SU(N)$ by a vev $\langle  (\Yfund,\Yfund) \rangle$, 
and also $[SU(N) \times SU(N),(\Yfund,\bar \Yfund) + (\bar \Yfund, \Yfund)]$ 
flows to a diagonal SQCD. 
Thus, none of the theories 
(\ref{pos1})   can be subprincipal. 
We conclude that  the  only possibility for a semisimple 
subprincipal stratum is that one of the projections, say $G_{(1)}$, 
be either 
$[Sp(2n),(2n+2) \Yfund]$ or $[SU(2),4 \Yfund]$. 
A reasoning similar to that leading to eq.(\ref{fla}) shows that if this 
is the case then 
the $G_{(2)}$ projection must contain an irrep $\rho_{\a}$ 
with a number of flavors $f_{\a} \geq d_{\a} -2$. All such theories, 
obtained by inspection of the tables in \cite{gn1,gn2},  
are listed in table  \ref{t1}.
It is a tedious but straightforward exercise to check that every one of the  
$24$ combinations $(i,j), i=1,2$ and $j=1-12$ flows to a nontrivial 
theory by Higgs mechanism, therefore none of them can be subprincipal.
\end{proof}

\setlength{\tabcolsep}{8mm}
\TABLE[h]{
\begin{tabular}{|r|c|c|}
\hline
 \multicolumn{3}{|c|}{{\sc Theories having an irrep $\rho_i$ with $f_i = d_i +2$}}\\  
\hline \hline
1 & $Sp(2n)$ & $(2n+2) \Yfund$ \\
2 & $SU(2)$ & $4 \Yfund$ \\ \hline \hline
 \multicolumn{3}{|c|}{{\sc Theories having an irrep $\rho_i$ with $f_i = d_i $}}\\  
\hline \hline
3 & $SU(N)$ & $N(\Yfund + \bar \Yfund)$ \\
4 & $Sp(4)$  & $\Yasymm + 4 \Yfund$ \\
\hline \hline
 \multicolumn{3}{|c|}{{\sc Theories having an irrep $\rho_i$ with $f_i = d_i - 1$}}\\  
\hline \hline
5 & $SU(N)$ & $\Yasymm + (N-1) \bar \Yfund + 3 \Yfund, \hspace{.4cm} N>4$\\
6 & $SU(4)$ & $\Yasymm + 3 \bar \Yfund + 3 \Yfund$\\
\hline \hline
 \multicolumn{3}{|c|}{{\sc Theories having an irrep $\rho_i$ with $f_i = d_i - 2$}}\\  
\hline \hline
7 & $SU(5)$ & $\Yasymm + 4 \bar \Yfund + 3 \Yfund$ \\
8 & $SU(4)$ & $2\Yasymm + 2 (\Yfund + \bar \Yfund)$ \\
9 & $SU(5)$ & $2 \Yasymm + \Yfund + 3 \bar \Yfund$ \\
10 & $SU(6)$ & $2 \Yasymm + 4 \bar \Yfund$ \\
11 & $Sp(6)$ & $\Yasymm + 4 \Yfund$ \\
12 & $Sp(4)$ & $2 \Yasymm + 2 \Yfund$ \\
\hline
\end{tabular}
\caption{Theories with a QMMS with $f_i \geq d_i -2$ flavors 
of matter in an irrep $\rho_i$ of  dimension $d_i$}
\label{t1}
}

The reader may think that this theorem implies that the classical
moduli space of a QMMS theory has a single subprincipal stratum.
This is not correct, a theory may have many $[SU(2),\Yfund + singlets]$ 
strata. This happens when different D-flat points break the 
gauge group to non conjugated $SU(2)$ subgroups.\\
As an example, consider the covariant QMMS theory 
$[SU(4), 3 \Yasymm + \Yfund + \bar \Yfund]$ \cite{gn1}. 
The D-flat point 
\begin{equation} \label{df1}
 A^1 = A^2 = 0 \hspace{.5cm} A^3 = \left( \begin{array}{rrrr} 
0&1&0&0\\ -1&0&0&0\\ 0&0&0&1\\ 0&0&-1&0 \end{array} \right), \hspace{.5cm}
Q= \left( \begin{array}{r} 0\\0\\0\\1 \end{array} \right), \hspace{.5cm} 
\bar Q = \left( \begin{array}{rrrr} 0&0&0&1 \end{array} \right)
\end{equation}
breaks $SU(4)$ to its subgroup
\begin{equation}
SU(2)_1 = \left\{ \left(\begin{array}{rr} g&0\\0&1 \end{array} \right), 
g \in SU(2) \right\}
\end{equation}
whereas the D-flat point $Q= 0,  \;\bar Q = 0$ and 
\begin{equation} \label{df2}
 A^1 =  \left( \begin{array}{rrrr} 
0&-1&0&0\\ 1&0&0&0\\ 0&0&0&-2\\ 0&0&2&0 \end{array} \right), 
\hspace{.5cm} 
A^2 =  \left( \begin{array}{rrrr} 
0&-2&0&0\\ 2&0&0&0\\ 0&0&0&-1\\ 0&0&1&0 \end{array} \right), 
\hspace{.5cm} 
A^3 = \left( \begin{array}{rrrr} 
0&0&1&0\\ 0&0&0&1\\ -1&0&0&0\\ 0&-1&0&0 \end{array} \right),
\end{equation}
breaks it down to the  diagonal 
\begin{equation}
SU(2)_2 = \left\{ \left(\begin{array}{cc} g&0\\0&g^{-1} \end{array} \right), 
g \in SU(2) \right\}
\end{equation}
The easiest way to see that these two $SU(2)$ subgroups are not conjugated 
is to notice that the eigenvalues of an element of 
$SU(2)_1$ are $(e^{i \a},e^{-i \a},1,1)$ whereas those of an $SU(2)_2$ 
element are $(e^{i \a},e^{-i \a},e^{i \a},e^{-i \a})$.
This theory has seven strata, ordered according to the diagram: 
\begin{equation}
\begin{array}{lclclclcl}
&& Sp(4) & - & SU(2) \times SU(2) & 
- & SU(2)_2 &  & \\
& \diagup & & & & \diagdown & & \diagdown&\\
SU(4) & - &SU(3)& \multicolumn{3}{c}{\frac{\hspace{3cm}}{}}&SU(2)_1&-& 1\\
\end{array}
\end{equation}
The possible Higgs flows are:
\begin{eqnarray} \nonumber
&&[SU(4), 3 \Yasymm + \Yfund + \bar \Yfund] 
\overset{ \langle \Yfund + \bar \Yfund \rangle}{\longrightarrow}
 [SU(3),3(\Yfund+\bar \Yfund )+{\mathbb I}]  
\overset{ \langle \Yfund + \bar \Yfund \rangle}{\longrightarrow}
 [SU(2)_1,4\Yfund + 6 {\mathbb I}] 
\overset{ \langle \Yfund  \rangle}{\longrightarrow}
[1, 11  {\mathbb I}] \; ,\\ \nonumber 
&&[SU(4), 3 \Yasymm + \Yfund + \bar \Yfund] 
\overset{\langle{\Yasymm} \rangle}{\longrightarrow}
 [Sp(4),2 \Yasymm+2 \Yfund+3 {\mathbb I}] 
\overset{ \langle \Yasymm \rangle}{\longrightarrow}
[SU(2) \times SU(2), 
(\Yfund,\Yfund) + 2(\Yfund,{\mathbb I}) + \\ \nonumber
&&2({\mathbb I},\Yfund) + 5 {\mathbb I}] 
\overset{ \langle (\Yfund,\Yfund)  \rangle}{\longrightarrow}
 [SU(2)_2,4\Yfund + 6 {\mathbb I}] 
\overset{ \langle \Yfund  \rangle}{\longrightarrow}
 [1, 11  {\mathbb I}]\; , \text{ and } \\ \nonumber 
&&[SU(4), 3 \Yasymm + \Yfund + \bar \Yfund] 
\overset{\langle{\Yasymm} \rangle}{\longrightarrow}
 [Sp(4),2 \Yasymm+2 \Yfund+3 {\mathbb I}] 
\overset{ \langle \Yasymm \rangle}{\longrightarrow}
[SU(2) \times SU(2), 
(\Yfund,\Yfund) + 2(\Yfund,{\mathbb I}) + \\ \nonumber
&&2({\mathbb I},\Yfund) + 5 {\mathbb I}] 
\overset{ \langle 2({\mathbb I},\Yfund)  \rangle}{\longrightarrow}
 [SU(2)_1,4\Yfund + 6 {\mathbb I}] 
\overset{ \langle \Yfund  \rangle}{\longrightarrow}
 [1, 11  {\mathbb I}].
\end{eqnarray}
We will come back to this example in Section \ref{teo}.

\section{Computing $\pi_j({\cal M}_{SQCD})$ in three different ways} \label{alg}

Supersymmetric SQCD with $N$ colors and flavors is the best known example 
of a theory with a QMMS. The elementary fields are 
the quarks $Q^{i\a}$ and $\bar Q_{j \b}$, 
the basic invariants are the mesons $M^i_j = Q^{i\a} \bar Q_{j \a}$ and 
baryons $B = \text{det } Q, \bar B = \text{det } \bar Q$. 
These  are subjected 
to the constraint 
\begin{equation} \label{qcdc}
\text{det } M - B \bar B = 0,
\end{equation}
which is quantum modified to \cite{sei,susy}
\begin{equation} \label{qcdq}
\text{det } M - B \bar B = \L^{2N}
\end{equation}
In the particular case $N=2$, (\ref{qcdq}) can be re written \cite{m}
\begin{equation} \label{qcdq2}
\vec z \cdot \vec z = \sum_{i=1}^6 {z_i}^2 = \L^4,
\end{equation}
where the components $z_i$ of the $SO(6)$ vector $\vec z$ 
are linearly related to  $(M,B,\bar B)$. As mentioned in the introduction, 
in this case the flavor group gets enlarged to 
$SU(4) \sim SO(6)$. This group  acts transitively on the 
QMMS, which can therefore be regarded as $SO(6,\c)/SO(5,\c) = (SU(4)/Sp(4))^c$.  
All this observations, made in \cite{m}, imply that, in the case $N=2$, 
$\m$ is homotopically equivalent to $S^5$, as explicitely shown 
by the deformation retraction (here $\vec z \in \c^6$ is written as 
$\vec{z} = \vec x+i \vec y$):
\begin{equation}
\phi(\vec x + i  \vec y,s)  =  \sqrt{\frac{\L^4+s^2 \vec y \cdot \vec y}
{\L^4+ \vec y \cdot \vec y}} 
\; \vec x + \, i \,  s \vec y , 
\hspace{.5cm}  0 \leq s \leq 1
\end{equation}
Since the quantum modification removes the origin and smoothes the
classical  moduli space, 
one may think, in view of Theorem \ref{teo1}, that $\m$ might be some sort 
of deformation of the principal stratum of $\mc$, and that we might obtain topological 
information  of $\m$ by looking at the principal stratum of $\mc$. 
Two color, two flavor 
SQCD is a good example  to show that this idea is wrong. 
The principal stratum of the classical moduli space of this theory is defined by 
\begin{equation}
\vec z \cdot \vec z = \sum_{i=1}^6 {z_i}^2 = 0, \;\;\; \vec z \neq \vec 0
\end{equation}
The deformation retraction
\begin{equation}
\phi(z,s) = \left[ \left( \sqrt{\frac{2}{\vec z \cdot {\vec z}^*}} \; -1 \right) s
 + 1 \right] z, 
\hspace{.5cm}  0 \leq s \leq 1
\end{equation}
shows that the principal stratum  
is homotopically equivalent to the Stiefel manifold \cite{steenrod}
\begin{equation}
V_{(6,2)} = \{ \vec x, \vec y \in {\mathbb R}^6 | \vec x \cdot \vec x = 
\vec y \cdot \vec y = 1, \vec x \cdot \vec y = 0 \}
\end{equation}
This set has many different possible interpretations, two of which are: 
(i) ordered sets $(\vec x, \vec y)$ of $2$ orthonormal 
vectors in ${\mathbb R}^6$, (ii) 
 bundle of unit tangent vectors of $S^5$ (here $\vec{y}$ is regarded 
as a unit vector, tangent at $\vec{x} \in S^5$). 
Homotopy groups of Stiefel manifolds can be found, e.g., in \cite{enc}, 
the topology of $V_{(6,2)}$ is completely different 
to that of $S^5$, 
there is no relation between $\m$ and the principal stratum of $\mc$.\\
In the computation of the homotopy type  of the moduli space 
(\ref{qcdq}) for $N>2$ in \cite{r}, new coordinates 
$(B=B_1+iB_2,\bar B=B_1-iB_2) $ are introduced such that  (\ref{qcdq}) gives 
$\text{det } M = \L^{2N} -{B_1}^2 -{B_2}^2$,  then it is shown that 
there is a retraction of this set onto the one defined by 
the same equation but with real $-\L^N \leq B_i \leq \L^N, i=1,2$, 
which  is a double suspension
  of the set $\{ M | \text{ det } M = 1\} 
= SL(N,\c) \sim SU(N)$. The homotopy groups of $\m$ can then be 
obtained from those of $SU(N)$ using Freudenthal's suspension theorem 
 ($\sim$ denotes same homotopy type, 
the definition of suspension and the statement of Freudenthal's theorem can 
be found in the appendix). \\
 As mentioned in the introduction, this idea cannot be 
extrapolated to other QMMS, because it strongly uses 
the form (\ref{qcdq}) of the SQCD constraint. 
An alternative approach is to use the 
results by Oka in   \cite{o} about  fibers of weighted homogeneous 
polynomials.  
A weighted homogeneous polynomial (w.h.p.) $p: \c^n \to \c$
is one that satisfies an 
equation of the form 
\begin{equation} \label{whp}
p(z^{w_1}x_1,..., z^{w_n}x_n) = z^d p(x_1,...,x_n)
\end{equation}
for some  positive integers $w_i$ called {\em weights}. 
Classical moduli spaces defined 
by a single constraint are of the form $p(x)=0$, with $p(x)$ 
 a w.h.p. The weight of $x_i$ is its  mass dimension, 
$d$ being the mass dimension of $p$. These sets are contractible, therefore 
trivial. This is easily seen from (\ref{whp}), if 
$p(x_1,...,x_n)=0$
then $p(z^{w_1}x_1,...,z^{w_n}x_n)=0$. 
A deformation retract of $\mc$ to the  point $x=0$ is then given by 
$(z^{w_1}x_1, \cdots, z^{w_n}x_n), z \in [0,1]$. 
If the classical moduli space $p(x)=0$ gets quantum modified and $p(x)$ 
is a flavor singlet,  the resulting {\em invariant} QMMS \cite{gn1} is the set 
\begin{equation}
p(x_1,...,x_n) = \L^d,
\end{equation}
If, instead,  $p(x)$ transforms non trivially under a flavor $U(1)$, 
we get a {\em covariant} type of QMMS 
\begin{equation}
p(x_1,...,x_n) = \L^{d-w_n}  x_n
\end{equation}
where the $U(1)$ charges of $p$ and $x_n$ agree, and 
the invariants have been properly numbered.
The scale of the theory is irrelevant, the moduli spaces 
defined by $\L$ and $\L'$ are  made diffeomorphic  by 
the map $x_i \to z^{w_i} x_i$, $z=\L' / \L$.\\
The derivative of (\ref{whp}) with respect to 
$z$, at $z=1$ 
\begin{equation}
\sum_{i=1}^s x_i w_i \partial_i p(x) = d p(\hp)
\end{equation}
shows that if $x$ is a critical point of $p$ ($\partial_ip(x)=0$) 
then $p(x)=0$. In particular, the fibers (level sets) 
$\{ x | p(x) = u \}$, $u \neq 0$ are smooth. 
Generically, the fiber over cero of a w.h.p. contains 
singular points, these are the well known singularities of 
$\mc$. 
Oka's theorem states that if  a w.h.p. splits
into  w.h.p's on different sets of variables
$p(x)=q(y)+r(z), x=(y,z)$ then, 
\begin{equation}
\{ x | p(x) = 1 \} \sim 
\{ y  | q(y) = 1 \}  * \{ z  | r(z) = 1 \}
\end{equation}
Here $A * B$ means the {\em join} of the sets $A$ and $B$ 
(see the appendix 
and \cite{gray}) 
which is obtained 
 by taking the disjoint union of the two spaces and 
connecting every point in $A$ to every point in $B$ by a line segment
(joining sets is an associative and commutative operation). 
The relation between the cohomology groups of $A * B$ to those of $A$ and $B$ 
is given, e.g.  in \cite{milnor,gray}. 
Oka's theorem is certainly well suited to study invariant QMMS's, 
since they are non singular fibers of w.h.p. 
As an example, consider the $N_c=N_f=2$ SQCD moduli space (\ref{qcdq2}).
Iterating Oka's theorem we arrive at 
\begin{equation}
\m_{(N=2)SQCD} \sim   \{ \text{2 points} \} * 
\{ \text{2 points} \}  *
 \{ \text{2 points} \}  * 
 \{ \text{2 points} \}  * \{ \text{2 points} \} * \{ \text{2 points} \}
\end{equation}
Using the associative property of joins 
we compute first  $\{  1,2 \}  *
 \{ a,b \} \sim S^1$ (see figure 1.a-b), then 
$(\{  1,2 \}  *  \{ a,b \} ) * \{ N,S \} \sim S^1 * \{ N,S \} \sim S^2$ 
(figure 1.c-d). More generally,  $S^{n-1}* \{ N,S \}  \sim S^n$, 
with $N$ and $S$ the poles of $S^n$, and $S^{n-1}$ its equator. 
Then $\m_{(N=2)SQCD} \sim S^5$ follows. \\

\begin{figure}[h]
\setlength{\unitlength}{0.45mm}
\begin{picture}(150,80)(0,-8) 
\put(0,30){\circle*{2}}
\put(4,28){1}
\put(56,30){\circle*{2}}
\put(60,28){2}
\put(30,56){\circle*{2}}
\put(34,54){b}
\put(34,4){a}
\put(30,4){\circle*{2}}
\put(0,-8){(a)  $\{ 1,2 \} \times  \{ a,b \}$}
%% 2 points * 2 points 
\put(84,30){\circle*{2}}
\put(88,28){1}
\put(140,30){\circle*{2}}
\put(144,28){2}
\put(114,56){\circle*{2}}
\put(118,54){b}
\put(118,4){a}
\put(114,4){\circle*{2}}
\put(75,-8){(b)  $\{ 1,2 \} *  \{ a,b \} \sim S^1$}
\qbezier(84,30)(94,48)(114,56)
\qbezier(84,30)(94,14)(114,4)
\qbezier(114,56)(124,40)(140,30)
\qbezier(140,30)(124,20)(114,4)
%% S^1 X 2 points 
\qbezier(170,30)(198,22)(226,30)
\qbezier(170,30)(198,38)(226,30)
\put(230,28){$S^1$}
\put(200,56){\circle*{2}}
\put(204,54){N}
\put(204,4){S}
\put(200,4){\circle*{2}}
\put(175,-8){(c)  $S^1 \times  \{ N,S \}$}
%% S^1 * 2 points 
\qbezier(260,30)(288,22)(316,30)
\qbezier(260,30)(288,38)(316,30)
\put(320,28){$S^1$}
\put(290,56){\circle*{2}}
\put(294,54){N}
\put(294,4){S}
\put(290,4){\circle*{2}}
\qbezier(290,56)(260,30)(290,4)
\qbezier(290,56)(230,30)(290,4)
\qbezier[60](290,56)(300,30)(290,4)
\qbezier(290,56)(320,30)(290,4)
\qbezier(290,56)(342,30)(290,4)
\put(250,-8){(d)  $S^1 *  \{ N,S \} \sim S^2$}
\end{picture}
\caption{The join of $S^{n-1}$ with a two point set gives $S^n$}
\end{figure}
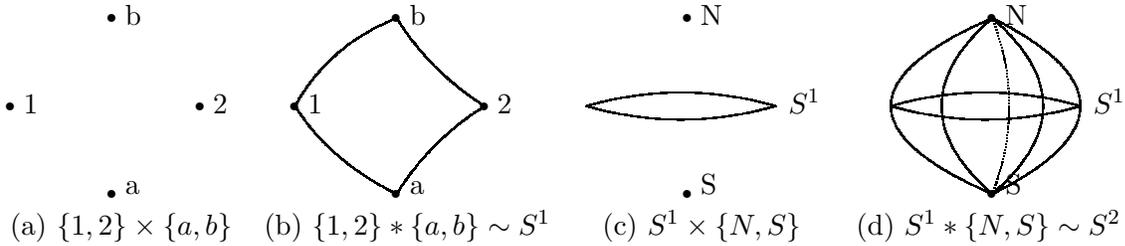

In the general case $N_c=N_f \geq 2$, 
the w.h.p. that defines $\m$,  $p(M,B,\bar B) = 
\text{ det } M - B \bar B$ is the sum of the polynomial on $N^2$ variables 
$q(M) = \text{ det } M$, whose non singular fiber 
$ \{ M |\text{ det } M = 1 \} = SL(N,\c) \sim SU(N)$, and the polynomial 
in two variables $r(B,\bar B) = B \bar B$, whose non zero fiber can be 
given as global coordinates $B \in \c - \{ 0 \} \sim S^1$. 
We conclude that
\begin{equation} \label{join}
\m_{SQCD} \sim SU(N) * S^1 = (SU(N) * \{ \text{2 points} \})  *
 \{ \text{2 points} \}  
\end{equation}
Since joining with a two point set gives the suspension (appendix A), 
(\ref{join}) is equivalent to the result in \cite{r} that 
$\m \sim $ double suspension of $SU(N)$. Back to the particular 
case $N=2$, we recover $\m \sim SU(2)*S^1 \sim S^3 * S^1 \sim S^5$.
In any case, (\ref{join}) implies $\m$ is 4-connected and has 
$\pi_5(M) = \mathbb Z (=H_5(\m, \mathbb Z))$, in view of Hurewicz theorem 
in the appendix. \\
Although Oka's theorem looks  very well adapted to the problem at hand, 
there is a problem: 
the {\em only} invariant   QMMS defined by  a  
polynomial constraint that can be usefully separated 
in polynomials of different variables seems to be  SQCD's. \\
In a recent paper by Dimca and Paunescu \cite{d}, an
alternative approach to study the topology of hypersurfaces defined by 
complex polynomial equations is given.  Their main result is 
the following:

\begin{theorem} [Dimca, Paunescu] \label{d&p} \cite{d} 
Let $f(x_1,...,x_n)$ be an arbitrary polynomial, $w_i > 0$ an 
arbitrary weight assignment to $x_i$, $f=f_d+f_e+ \cdots f_0$ 
the decomposition of $f$ in weighted homogeneous components 
of degrees $d>e> \cdots >0$. Let 
\begin{equation} \label{dim}
S = \{x \in \c^n | \partial_i f_d = 0, f_e=0 \}
\end{equation}
Any fiber $f(x)=c$ is $q-$connected, with $q=n-2- \text{ dim } S$
\end{theorem}

Let us see what this theorem says for SQCD with $N$ colors and flavors, 
eq(\ref{qcdq}). 
Assigning the usual weights $w_M = 2, w_B = w_{\bar B} = N$ we get  
$f_d = \text{ det } M -B \bar B, f_e=0$ and $S = \{ (M,B,\bar B) \in \c^{N^2+ 2}  |
B=\bar B = 0, \; \text{ rank } M  \leq N-2 \}$. 
We are interested in the non singular fiber of a polynomial $f$ 
in $n=N^2+2$ variables. $S$ is the algebraic set of $N$ by $N$ matrices 
of rank $\leq N-2$, its dimension 
is $N^2 - 4$ \cite{ls}, then 
(\ref{dim}) says that both $\m_{SQCD}$  
(and certainly ${\mc}_{SQCD}$, which is another fiber of the same 
polynomial) is $N^2+2-2-(N^2-4) = 4-connected$. 
From Hurewicz theorem (appendix A) we know that $H_5(\m_{SQCD},{\mathbb Z}) 
\simeq \pi_5(\m)$. However Theorem \ref{d&p}  does not tell us  what this group is 
(had we obtained $q=5$ then  also $H_5$ would be trivial and the 
theory would not admit Wess-Zumino terms.)  In the SQCD case, 
we know that Theorem \ref{d&p}
has given us the best possible estimate for the connectedness of $\m$, 
but this may not always happen. Note the following subtlety \cite{d}, 
the theorem works for any 
(positive) weight assignment to the $x_i$'s, and  a given 
weight assignment  is likely to 
give better estimates than others if it makes most monomials 
belong to $f_d$ (as happens with the natural weight assignment 
in both types of QMMS constraints.)  
Back to SQCD, if instead of the natural weight assignment 
given by the mass dimension of the operators we use
$w_M =  w_B = w_{\bar B} = 1$,  then $f_d = \text{ det } M, 
f_e = B \bar B$, and $S = \{ (M,B,\bar B) \in \c^{N^2+ 2}  |
B\bar B = 0, \; \text{ rank } M  \leq N-2 \}$. With this weight assignment 
we obtain dim $S = N^2 -3$, then $q=3$. Although it is certainly correct that 
$\m_{SQCD}$ is $3-$connected, this is not the 
best estimate. \\
Although Theorem \ref{d&p} gives only partial information about 
the topology
of $\m$,  it has the advantage that 
can be applied to any QMMS, since it does not make any assumptions on
 the polynomial that defines $\m$. 
This makes it very powerful, especially because there is a way to 
 estimate dim $S$, without even knowing the polynomial that defines it!
This is the subject of the next section.

\section{Computing $\pi_j(\m)$ for all hypersurface-like QMMS} \label{teo}

\begin{theorem} \label{d2} If $\m$ is a quantum modified moduli space defined by a single 
constraint then $\m$ is 4-connected, i.e, it is connected and 
has trivial homotopy groups $\pi_j(\m)$ for $j=1,2,3,4$.
\end{theorem}

\begin{proof}
Consider first the {\em invariant} QMMS \cite{gn1,gn2}
\begin{equation} \label{iqmms}
\m = \{ \hp \in \c^n | p(\hp) = \L^d \} \subset \c^n,
\end{equation}
$n$ the number of basic invariants.
If $d_i$ is the mass dimension of $x_i$ and $d$ 
the mass dimension of $p(\hp)$, then 
\begin{equation}\label{wh}
p(z^{d_1} \hp_1, z^{d_2} \hp_2,\cdots,z^{d_k} \hp_k) = 
z^d p(\hp_1,\hp_2,\cdots,\hp_k),
\end{equation}
showing that $p$ is a weighted homogeneous polynomial. 
The derivative of (\ref{wh}) with respect to $z$  at $z=1$ gives 
\begin{equation}
\sum_{i=1}^s \hp_i d_i \partial_i p(\hp) = d p(\hp)
\end{equation}
According to Theorem \ref{d&p}, the set (\ref{iqmms}) is $q-$connected, 
with 
\begin{eqnarray} \label{q1}
 q &=& n-2-dim(S), \\
 S &=&\{ \hp \in \c^n | \partial_i p(\hp) = 0 \} \label{q1b}
\end{eqnarray} 
From equation (\ref{wh}) follows that $\partial_i p(\hp) = 0$ implies 
$p(\hp)=0$, then $\hp \in \mc$. Moreover, $\hp$ is 
a {\em singular point} of $\mc$, since $\partial_i p(\hp) = 0$. 
According to Theorem 1.3  $\hp$  lies outside 
the principal stratum, or, equivalently, 
in  the union of the closures of 
the subprincipal strata, since this set contains all non-principal strata 
(see Theorem  1.2 and the example given in eqns. (\ref{ex}), (\ref{ex1}) and 
(\ref{ex2})). Thus,  the  dimension of $S$ is smaller
than  or equal to 
that of the highest dimensional subprincipal stratum.
Theorem \ref{d1}  says that the only possible kind of subprincipal stratum is 
$[SU(2),4 \Yfund + s {\mathbb I}]$. We conclude that 
\begin{equation} \label{bound}
\text{dim } S \leq \text{dim } \Sigma_{(SU(2))}.
\end{equation}
The dimension of the $SU(2)$ strata is given by the number $s$ of singlets 
(Theorem 1.4). If $\phi \in \c^k$ is a D-flat point that breaks 
$G$ to $SU(2)$, then $\c^k$ splits into the $SU(2)$ invariant 
subspaces $Lie (G) \phi \oplus 4 \Yfund \oplus s $ singlets.
Since the dimension of the $G\phi$ orbit is dim $Lie(G) \phi = 
\text{dim } G -\text{dim } SU(2) = \text{dim } G -3$, it follows 
that 
\begin{eqnarray} \nonumber
\text{dim } \Sigma_{(SU(2))} = 
s &=& k - \text{dim } Lie(G) \phi - \text{ dim } 4 \Yfund \\ \nonumber
&=& k - \text{ dim } G - 5 \\ \nonumber
&=& \text{dim } \mc - 5 \\ \label{q2}
&=& n - 6
\end{eqnarray}
From (\ref{q1}), (\ref{bound}) and (\ref{q2})
\begin{equation} \label{q}
q = n - 2 - \text{ dim } S \geq  n - 2 - \text{dim } \Sigma_{(SU(2))} = 4, 
\end{equation}
from where we conclude $\m$ is $4$-connected.\\
Consider now the covariant QMMS
\begin{equation} \label{cqmms}
\m = \{ \hp \in \c^n | p(\hp) - \L^{d-d_n} \hp_n =0 \} \subset \c^n,
\end{equation}
In this case (\ref{q1b}) has to be replaced with 
\begin{equation}
S = \{ \hp \in \c^n | \partial_i p(\hp) = 0 \text{ and } \hp_n=0 \} \label{s4}
\end{equation}
which is the set of critical points of the classical moduli 
space with $x_n=0$. Once again, this set is 
included in the $SU(2)$ strata, and (\ref{q}) follows.
\end{proof}
 At first sight, the extra condition $x_n=0$ 
may suggest that we could get a better by one estimate of 
 the dimension of $S$. This 
 would imply the $5-$connectedness of the covariant QMMS, and, 
in view of 
of Hurewicz theorem, that 
these theories do not admit 
Wess-Zumino-Witten terms. However, this analysis is wrong. 
Computing the dimension of an algebraic 
set like $S$ in Theorem \ref{d&p} is a subtle issue (see, e.g., \cite{cls}). 
The algebraic set $S$ decomposes into {\em irreducible components}, 
$S = S_1 \cup S_2 \cup \cdots \cup S_p$ ($S_j$ {\em irreducible} means
that it is not the proper union of two algebraic sets),  
the dimension of $S$ 
being  the maximum dimension of an irreducible component. 
In an irreducible set $X$, the dimension 
of the tangent space may change from point to point, the dimension 
of $X$ is the minimal dimension of a tangent space. 
As an example, consider again the covariant 
QMMS theory $[SU(4), 3 \Yasymm+ \Yfund + \bar \Yfund]$ from the 
end of section \ref{strat}. The basic invariants are \cite{gn1}
\begin{multline} \label{su4invs}
M = Q^{\a} \bar Q_{\a}, \; B_i =  A^{j \a \a_1} A^{k \a_2 \a_3}Q^{\a_4} 
\bar Q_{\a} \e_{\a_1 \a_2 \a_3 \a_4} \e_{ijk}, \;
 P = A^{i \a \a_1} A^{j \b \a_2}
\bar Q_{\a} \bar Q_{\b} A^{k \a_3 \a_4}\e_{\a_1 \a_2 \a_3 \a_4} \e_{ijk},\\
S^{ij} = A^{i \a_1 \a_2} A^{j \a_3 \a_4} 
\e_{\a_1 \a_2 \a_3 \a_4},\;
R =  A^{i \a_1 \a_2} A^{j \a_3 \b_1} A^{k \b_2 \b_3} Q^{\a_4} Q^{\b_4}
\e_{\a_1 \a_2 \a_3 \a_4}\e_{\b_1 \b_2 \b_3 \b_4}   \e_{ijk}
\end{multline}
The classical constraints is \cite{gn1}
\begin{equation} \label{csu4}
p(M,B,S,P,R) \equiv M^2 \text{ det } S + c_1 S^{ij}B_i B_j + c_2 PR = 0,
\end{equation}
with $c_1$ and $c_2$ nonzero constants. 
The QMMS is defined by 
\begin{equation} \label{csu4q}
M^2 \text{ det } S + c_1 S^{ij}B_i B_j + c_2 PR = \L^8 M.
\end{equation}
The set $\Sigma$ of critical points of $p(M,B,S,P,R)$ has 
two irreducible components, defined by 
\begin{equation} \label{s1}
\Sigma_1: (M,B,S,P,R) \text{ such that }
\begin{cases} 
\text{ det } S = 0\\
M^2 \text{ cof } (S)_{ij} + c_1 B_i B_j = 0\\
S^{ij}B_j = 0\\
P=R=0
\end{cases}
\end{equation}
\begin{equation} \label{s2}
\Sigma_2: (M,B,S,P,R) \text{ such that } \begin{cases} 
M  = 0\\
M^2 \text{ cof } (S)_{ij} + c_1 B_i B_j = 0\\
S^{ij}B_j = 0\\
P=R=0
\end{cases}
\end{equation} 
The set $S$ in Theorem \ref{d&p} is 
$S =\{ (M,B,S,P,R) \in \Sigma | M=0 \} = \Sigma_2$. 
Note that (\ref{s2})  is equivalent to $M=B_i=P=R=0$, 
$S^{ij}$ an arbitrary (symmetric) tensor, so $\Sigma_2$  has dimension six.
In (\ref{s1}), $S^{ij}$ and $M$ determine $B_i$ from the second 
equation, which satisfies the third equation. So we can freely 
choose a symmetric, singular $S^{ij}$ and $M$, meaning that 
dim $\Sigma_1$ also equals six.
From Theorem \ref{d&p} $\m$ is $q-$connected with 
$q=12-2- \text{ dim } S = 4$. 
The extra condition, 
$M=0$ does not make dim $S < $ dim $\Sigma$, 
instead, it projects   onto one of the two 
six dimensional irreducible components of the set $\Sigma$ of singular points of $\mc$.
As a result, the estimate from Theorem \ref{d&p} is that $\m$ is $4-$connected,
not $5-$connected  as one might  have first thought. Note 
from (\ref{df1}), (\ref{df2}), (\ref{s1}) and  (\ref{s2}) that 
$\Sigma_i = \overline{\Sigma_{(SU(2)_i)}}, i=1,2$, the fact that there are two 
irreducible components of $\Sigma$ is related to the fact that there are  
two $SU(2)$ strata in this theory.
Although this example shows that 
Theorem \ref{d2} gives the best possible estimate for the connectedness 
of $\m$, the  possibility of having a $5$-connected QMMS is
not ruled out. It may well happen that the extra condition 
$x_n=0$ in (\ref{s4}) does lower the dimension of $S$ for a particular
 covariant QMMS. Also, 
since the converse of Theorem 1.3 is not true, i.e., smooth
points with enhanced gauge symmetry can be found in $\mc$  \cite{t}, 
it is possible to have a strict inequality in (\ref{bound}), 
then also in (\ref{q}). Thus,  
even  invariant QMMS may be $5-$connected.
Since this requires a case by case verification, 
we have applied Theorem \ref{d&p} to a sample  of QMMS defined 
by a single constraint, available in the literature. All cases 
turned out to be $4-$connected 
with the  
exception of the theory $SU(6)$ with $\Ythreea + \Yasymm + 2\bar \Yfund$, 
which, according to Theorem \ref{d&p} and the constraint equation given 
in \cite{gn1}, has a $5-$connected quantum moduli space, 
and therefore a trivial $H_5$ (note that no WZ term is required to match 
       the broken $SU(2)$ anomalies.) We have verified that the $13$ invariants 
given in eqns (A.22)-(A.30) in \cite{gn1} for this theory is a complete set of 
basic invariants 
up to degree seven, and 
that the constraint equation in \cite{gn1}  (with minor irrelevant 
changes in some coefficients) reduces 
 to a classical constraint, and defines a smooth twelve dimensional 
variety \footnote{We have not ruled out  
the possibility of having  basic (i.e., algebraically independent from 
those of lower degree) invariants of higher degree. These  should 
come together with additional constraints,   to ensure the condition
dim $\mc = 12$.}.
$SU(6)$ with $\Ythreea + \Yasymm + 2 \bar \Yfund$ 
is then an example of a theory with a QMMS that does not support 
Wess-Zumino terms, in spite of flowing, as every other  QMMS theory 
does, 
to  $SU(2)$ with $4 \Yfund$, a theory with Wess-Zumino terms in 
its effective action.

\section{Conclusions} \label{conc}
The stratification of the classical moduli space  $\mc$ of a 
supersymmetric gauge theory with a quantum modified moduli space $\m$ 
plays an unexpected role in the determination of  relevant 
topological aspects of  $\m$. In particular, the fact that 
these theories have an $[SU(2), 4 \Yfund]$ stratum implies that 
$\m$ is connected, simply connected, and also has trivial 
$\pi_j(\m)$ for $j=2,3,4$. As a consequence,  $\m$ 
does not support vortexes or skyrmions, these configuration can 
``unwind'' because  $\pi_2(\m)$ and  $\pi_3(M)$ are trivial. 
A trivial  $\pi_4(\m)$ is one of the necessary conditions 
to construct a 
Wess-Zumino-Witten functional on $\m$, the other requirement being 
a non trivial $H_5(\m)$, or,  equivalently (in view 
of Hurewickz theorem), a non trivial $\pi_5(\m)$.
Testing this last condition seems to  require a case by case analysis,  
together with  the application of new  
approaches that we are currently developing.

\acknowledgments
 I am grateful to  Reimundo Heluani for many helpful 
discussions, Laurent Paunescu and Alexandru Dimca 
for clarifying  aspects related to the  
application of their results. This work was supported by 
Conicet and Secyt-UNC.

\appendix  \section{Selected algebraic topology facts}

$\pi_o(X)$ is defined to be the  set of 
path connected components of $X$ (\cite{bott}, pp.206-), it 
does not have a group structure.  
$X$ is said to be $n-$connected if  $\pi_j(X)$ is trivial for $j=0,1,...,n$. \\
{\sc Hurewicz theorem:} (\cite{bott}, p.225) 
 Let $X$ be a simply connected, path connected CW complex. Then the 
first non trivial homotopy and homology occur in the same dimension 
and are equal, i.e., given a positive integer 
$n \geq 2$, if $\pi_q(X)=0$ for $1 \leq q < n$, then $H_q(X)=0$ for 
$1 \leq q < n$ and $H_n(X) = \pi_n(X)$.\\
Every manifold has the homotopy type of a  CW complex 
(\cite{bott},p.220, \cite{mm}, p.36),  also $H_q$ and $H^q$ are 
isomorphic groups, then we can 
re-state Hurewicz theorem as follows: if a manifold $X$
is $(n-1)$ connected then $H^q(X)$ is trivial if $q<n$, 
whereas $H^n(X) = \pi_n(X)$.\\
{\sc Join:} (\cite{gray}, p.334)
Given two topological spaces $X$ and $Y$, their {\em join,}
 denoted by $X * Y,$ is defined to be the quotient space
\[
X* Y := X\times[0,1]\times Y/\sim,
\]
where the equivalence relation $\sim$ is generated by
\begin{eqnarray*}
(x,0,y_1)& \sim (x,0,y_2) &\text{for any}\, x\in X,\, y_1,y_2\in Y,\, \text{and}\\
(x_1,1,y)& \sim (x_2,1,y) &\text{for any}\, y\in Y,\, x_1,x_2\in X.
\end{eqnarray*}
Intuitively, $X * Y$ is formed by taking the disjoint union of the two spaces and 
attaching a line segment joining every point in $X$ to every point in $Y.$\\
{\sc Suspension:} 
Given a topological space $X,$ the {\em suspension} of $X,$ often denoted by
 $SX,$ is defined to be the quotient space $X\times[0,1]/\sim,$ 
where $(x,0)\sim(y,0)$ and $(x,1)\sim(y,1)$ for any $x, y\in X.$
Note that $SX$ is homeomorphic to the join
 $X * S^0,$ where $S^0$ is a discrete space with two points, so $X * S^0$
can be taken as an alternative definition of $S X$.\\
{\sc Freudenthal's suspension theorem:}  if $X$ is $(n-1)-$connected, 
$\pi_q(X) \simeq \pi_{q+1}(SX)$ for $q \leq 2n-2$ and also there is an 
onto homomorphism $h:\pi_{2n-1}(X) \to \pi_{2n}(SX)$ (\cite{gray} 
p145 and p.135)

\end{document}